\PassOptionsToPackage{table}{xcolor}
\documentclass[sigconf,nonacm]{acmart}

\usepackage{booktabs}
\definecolor{agentblue}{HTML}{1E6091}
\definecolor{humanred}{HTML}{E97266}
\definecolor{groupgray}{HTML}{F0F0F0}
\newcommand{\eff}[2]{\textcolor{agentblue}{\rule[0.5pt]{#1}{6pt}}%
  \textcolor{humanred}{\rule[0.5pt]{#2}{6pt}}}

\begin{document}

\title{Test-Driven, AI-Assisted Learning: Replacing Lectures with Weekly Closed-Book Tests}

\author{Jin-Guo Liu}
\email{jinguoliu@hkust-gz.edu.cn}
\affiliation{%
  \department{Thrust of Advanced Materials, Function Hub}
  \institution{The Hong Kong University of Science and Technology (Guangzhou)}
  \city{Guangzhou}
  \country{China}
}

\author{Shang-Qi Lu}
\email{shangqilu@hkust-gz.edu.cn}
\affiliation{%
  \department{Thrust of Data Science and Analytics, Information Hub}
  \institution{The Hong Kong University of Science and Technology (Guangzhou)}
  \city{Guangzhou}
  \country{China}
}

\author{Xin-Ran Shi}
\affiliation{%
  \department{Thrust of Data Science and Analytics, Information Hub}
  \institution{The Hong Kong University of Science and Technology (Guangzhou)}
  \city{Guangzhou}
  \country{China}
}

\author{Long-Li Zheng}
\affiliation{%
  \department{Thrust of Advanced Materials, Function Hub}
  \institution{The Hong Kong University of Science and Technology (Guangzhou)}
  \city{Guangzhou}
  \country{China}
}

\author{Wei Wang}
\email{weiwcs@hkust-gz.edu.cn}
\affiliation{%
  \department{Thrust of Data Science and Analytics, Information Hub}
  \institution{The Hong Kong University of Science and Technology (Guangzhou)}
  \city{Guangzhou}
  \country{China}
}

\renewcommand{\shortauthors}{Liu et al.}

\begin{abstract}
This paper is an experience report on a 13-week \emph{Test-Driven,
AI-Assisted} (TDAA) redesign of DSAA 3071, Theory of Computation, an
upper-level course at the Hong Kong University of Science and Technology
(Guangzhou). The design is simple: the course replaces lectures with self-directed,
AI-assisted learning, and frequent, independently completed tests create a
high-frequency quality gate. AI agents help the instructor prepare
the learning path, course website, tests, grading workflow, and repairs.
Two conditions made this strict gate workable. Students needed a visible
preparation path of learning sheets and aligned validation practice, so the
closed-book tests felt fair rather than arbitrary. The instructor needed an
AI-assisted materials harness---a version-controlled agent workspace---so that
weekly drafting, review, test production, and grading could scale with human
oversight. Evidence from a
student survey ($N=18$), weekly scores, and the project's git history
suggests that students treated the tests as useful accountability and that
the harness made frequent closed-book testing operational. The evidence is
limited to one small, proof-heavy course without a control group. The
contribution is therefore a reusable design pattern: high-frequency tests
preserve individual accountability, while AI agents make material production and
marking scalable. We release the harness as a public starter template so
that other instructors can reproduce it.
\end{abstract}



\maketitle

\section{Introduction}
\label{sec:intro}

Lectures can make learning look more active than it really is. A teacher may
explain clearly, students may sit quietly, and the class may still leave many
students unable to solve problems on their own. In the worst case the
classroom becomes staged engagement: the teacher performs explanation,
students perform attention, and neither side has strong evidence that students
can use the ideas. This challenge predates generative AI. Active learning,
peer instruction, and flipped classrooms all point to the same lesson: students
usually learn more when they have to use ideas rather than only hear
them~\cite{freeman2014active,crouch2001peer,bishop2013flipped,lage2000inverting}.

Generative AI makes this old problem more urgent. Tools such as ChatGPT can
answer questions, explain concepts, and produce text, so they may lower the
cost of the repeated, individualized feedback that Bloom's two-sigma problem
made famous~\cite{bloom1984}. Recent education work therefore asks teachers to
design new learning opportunities instead of only allowing or banning
AI~\cite{mollick2024instructor}, and current assessment reviews argue that
traditional assessment must be rethought in GenAI
environments~\cite{weng2024genaiAssessment}. But AI also weakens a familiar
signal of learning: if students use it for homework or take-home projects, the
submitted work may no longer show what the student personally
understands~\cite{cotton2024chatting}.

Broad access to capable LLMs changes what an educator is for. When any student
can get an explanation on demand, the teacher's scarce contribution shifts from
delivering lectures toward designing the learning path and verifying mastery
under controlled conditions. A course therefore needs an accountability
mechanism that is both realistic and serious. Flexible student AI use is already
part of the learning
environment, with or without a course design. The contribution cannot simply be
letting students use AI. The better question is: how can a course give
students a clear path to independent performance, check that performance often
and seriously, and keep the instructor-side workload feasible?

\textbf{The TDAA design.} Our answer is \emph{test-driven, AI-assisted}
learning (TDAA), built on two coupled parts. The first is a shift from
teaching to evaluation. The course abandons traditional lecturing and lets
students learn by interacting with AI and the prepared materials at their own
pace. A frequent, independently completed (closed-book) test then makes that
self-directed learning individually accountable. Students demonstrate unaided
understanding weekly rather than waiting for a final exam. Frequent independent
testing is not only an assessment device but also a study aid. Paired with
aligned practice, it gives students a concrete target and a predictable signal
of where they stand. Section~\ref{sec:background} reviews the learning-science
evidence. The second part
is AI-assisted course production: agents help build the learning path, produce
learning materials and the course website, generate and review tests, mark
papers, and repair mistakes under human approval. \emph{The pairing is the
contribution.} Without the agent layer, weekly aligned testing is
operationally infeasible for a single instructor; without the test gate,
AI-assisted preparation offers no reliable check on whether students can
perform unaided.

\textbf{Reflection questions.} As an experience report, this paper asks three
questions of one real run: (RQ1) Did students accept a lecture-free
design, in which self-directed AI-assisted study sits behind a strict weekly
closed-book gate, when given an explicit, aligned path to prepare? (RQ2) Was
weekly aligned testing operationally feasible for a single instructor using a
human-reviewed AI harness? (RQ3) Where did the design strain, and what should
the next iteration fix? The first implementation was DSAA 3071, Theory of
Computation, at the Hong Kong University of Science and Technology (Guangzhou)
(HKUST(GZ)) in Spring 2026. Each week opened with a short closed-book test on the previous week's
material, then moved to self-paced study with instructor- and TA-supported use
of the prepared materials.

\section{Background and related work}
\label{sec:background}

\textbf{Testing as a learning mechanism.} Frequent low-cost testing is
supported by retrieval-practice research: taking a test improves later
retention~\cite{roediger2006testenhanced}, and practice testing reliably
improves later performance in meta-analytic
reviews~\cite{adesope2017practiceTesting}. TDAA uses this mechanism but raises
the stakes to weekly graded closed-book tests, so it pairs each test with
aligned preparation following constructive
alignment~\cite{biggs1996constructiveAlignment}.

\textbf{Self-paced study needs structure.} TDAA removes the lecture and asks
students to use AI for the just-in-time explanation a lecture would have given,
pushing flipped- and active-learning logic to its limit. Prior self-paced and
mastery-learning implementations show that without enough checkpoints,
incentives, and self-regulation support, students may procrastinate or use
learning tools superficially~\cite{campbell2019selfPacedMasteryCS1,devore2017selfPacedTutorials}.
TDAA's weekly gate is exactly such a checkpoint.

\textbf{GenAI and assessment.} GenAI weakens take-home work as an
individual-learning signal and is pushing assessment redesign~\cite{cotton2024chatting,weng2024genaiAssessment,mollick2024instructor}.
Broader cautions concern critical thinking, fact-checking, and the need for
human oversight~\cite{kasneci2023chatgpt}.

\textbf{LLM agents and AI grading.} LLM-based agents are commonly described as
systems that combine model reasoning with planning, memory, and
action~\cite{wang2024llmAgents}. Recent work reports that LLMs can support
criterion-based grading when detailed criteria are supplied, but that
handwriting/OCR errors remain a real risk for paper-and-pencil
assessment~\cite{zhang2024llmGrading,tan2024handwritingErrors}.

\textbf{Scaffolding hard topics.} Proof-heavy material is cognitively demanding,
and cognitive-load and proof-education work motivates worked examples and
scaffolded proof construction~\cite{gog2010cognitiveLoad,poulsen2023proofBlocks}.

\section{The TDAA learning model}
\label{sec:tdaa-learning}

\begin{figure*}[t]
  \centering
  \includegraphics[width=0.9\textwidth]{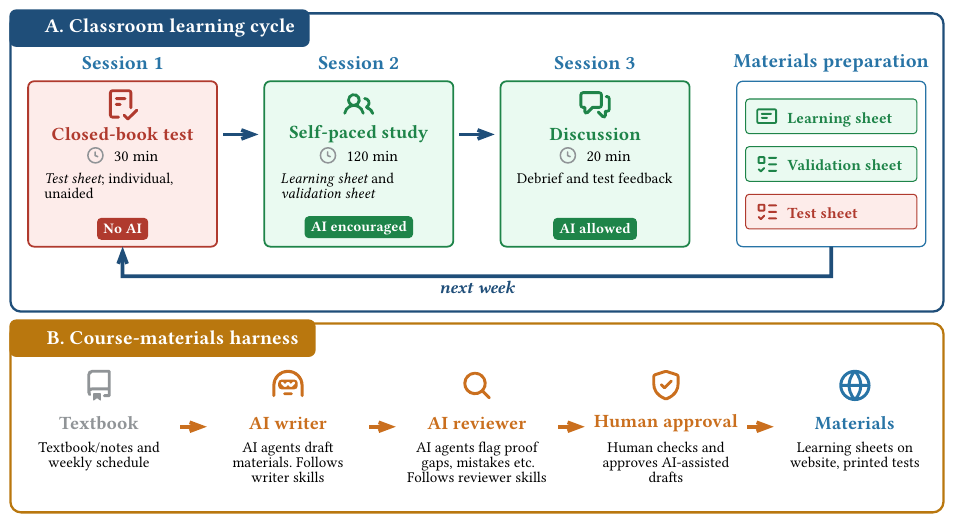}
  \caption{The TDAA learning model. Panel A: the weekly classroom cycle, where
  guided preparation feeds a strict closed-book quality gate. Panel B: the
  course-materials harness that makes frequent preparation, testing, review,
  and grading feasible.}
  \label{fig:overview}
\end{figure*}

\textbf{Model overview.} TDAA is two coupled cycles
(Figure~\ref{fig:overview}): in the \emph{student cycle}, students prepare with
guided materials and then face a weekly closed-book gate; in the
\emph{production cycle}, an instructor-side harness produces those materials and
grades the gate. The gate is strict by design; the preparation cycle is what
makes that strictness fair. We explain each below.

\textbf{Student cycle.} TDAA's classroom has no lecture: students learn by
working the prepared materials and questioning AI at their own pace, and a
weekly closed-book gate verifies they can perform unaided. The \emph{hard
quality gate} (Panel~A, leftmost card) is a closed-book test: no AI, notes,
books, or internet. High frequency makes
it powerful but also risky. Frequent tests alone can feel punitive, so TDAA
makes the gate acceptable by giving students a clear home base for
preparation. Three artifacts anchor that base: a \emph{learning sheet} naming
the week's ideas, definitions, notation, and examples; a \emph{validation
sheet} of practice problems; and the \emph{closed-book test} itself, written in
advance against the same scope. Weekly tests can also be capped (here, at 100
of 130 available points) so that full credit does not require a perfect paper.
Students reduce test pressure by studying the learning sheet, working the
validation problems, and fixing confusion before the gate. Without that
anchor, students drift to outside notes, videos, or AI answers whose notation
and ordering may not match the textbook.

\textbf{Production cycle.} Panel~B addresses the scalability problem: who
produces and marks so many test bundles? The answer is a three-stage pipeline
anchored to the course textbook: an AI writer drafts the material, an AI
reviewer challenges it, and a human approver makes the final call before
anything reaches students. AI operates only on the production side, drafting
learning sheets, generating and reviewing tests, and assisting with
marking. It never acts at the closed-book gate, and never without human sign-off.

\section{The course-materials harness}
\label{sec:harness}

The operational challenge was not only volume but reliability on a weekly
clock. Each weekly bundle had to agree internally on definitions, notation,
scope, and difficulty. Post-test marking had to be correct and fast enough to
return feedback before the next learning cycle. Three files per week across
twelve content weeks, plus post-test grading, yields 36 aligned
documents, a volume no instructor can keep consistent by hand. The production
cycle of Figure~\ref{fig:overview} therefore became a course-materials
\emph{harness}: a version-controlled workspace where AI agents do the repetitive
work. They draft, cross-check, build, write grading rubrics, and mark first
passes. The instructor reviews, repairs, and approves. Like a software test
harness, its job is less to do the work than to hold that work to a fixed
standard.

An \emph{AI agent} here is a program that wraps a large language model in a
tool-using loop. The model reads project files, calls tools
(\texttt{Read}, \texttt{Edit}, \texttt{Write}, \texttt{Bash}, \texttt{Grep},
\texttt{Glob}) to act on the repository, sees each result, and decides the next
step. Contemporary coding agents all follow this
pattern~\cite{anthropic2025claude,openai2026codex,cursor2026}. The tool layer
was deliberately ordinary: most generated materials were editable text-based
source files that could be built into PDFs or
webpages~\cite{typst}; Git preserved change history~\cite{git}. The important
requirement is not a particular product, but a version-controlled text
workspace where AI output is inspectable, rebuildable, and constrained by named
skills.

Following the harness structure of~\cite{liu2026predharness}, four pieces
surrounded the agent: (i) a \emph{project specification} (a
\texttt{CLAUDE.md} file for Claude Code, or \texttt{AGENTS.md} for other coding
agents) loaded automatically at the start of every session, naming the textbook,
weekly schedule, file layout, and writing conventions; (ii) a \emph{knowledge base} of textbook
content (Sipser's \emph{Introduction to the Theory of
Computation}~\cite{sipser2013introduction} converted to
Markdown),\footnote{The textbook was converted from a licensed copy and used
only as a private, instructor-side reference, not part of the public release.
Generated materials re-express concepts and cite the textbook rather than
reproducing its text, figures, or exercises.} the weekly schedule, and prior
finalized materials; (iii) a set of \emph{tools} the agent
invokes during a task, such as \texttt{Grep} over the textbook to look up a
definition, web search to cross-check a proof claim, image reading to mark
scanned tests, a document compiler to verify a draft builds, and Git to record
every change; and (iv) a set of \emph{skills}, named instructions for repeated
tasks such as drafting a learning sheet, reviewing it, or grading a test.

Skills are the orchestration layer: each composes the project spec, knowledge
base, and tools into an ordered checklist the agent executes faithfully. This
prevents the drift and omissions that arise when an agent plans from scratch.
Following~\cite{liu2026predharness} we distinguish \emph{automation skills},
which run end-to-end, from \emph{advisor skills}, which pause for instructor
judgment. In practice the split is a gradient, visible in
Table~\ref{tab:skills}'s effort bars. Drafting and review run with almost no
human time. The \texttt{revise} skill is mostly human. Grading sits between: it
automates routine marks but escalates uncertain ones.
\begin{table}[t]
  \caption{Named AI-assisted skills used to produce and review weekly
  materials, grouped by role. The effort bar estimates relative effort:
  \textcolor{agentblue}{\rule[0.5pt]{3mm}{6pt}}~agent,
  \textcolor{humanred}{\rule[0.5pt]{3mm}{6pt}}~human.}
  \label{tab:skills}
  \footnotesize
  \setlength{\tabcolsep}{4pt}
  \begin{tabular}{@{}>{\raggedright\arraybackslash}p{0.22\columnwidth}>{\raggedright\arraybackslash}p{0.52\columnwidth}c@{}}
    \toprule
    \textbf{Skill} & \textbf{What it does} & \textbf{Effort} \\
    \midrule
    \multicolumn{3}{@{}l}{\cellcolor{groupgray}\textit{Setup}}\\
    \texttt{bootstrap} & Set up a new course from a forked template; confirm metadata, write config and draft schedule. & \eff{9.75mm}{5.25mm}\\
    \multicolumn{3}{@{}l}{\cellcolor{groupgray}\textit{Authoring (AI writer)}}\\
    \texttt{generate\allowbreak-week} & Generate a full week (learning sheet + tests) via writer-vs-reviewer debate. & \eff{13.5mm}{1.5mm}\\
    \texttt{write\allowbreak-learning\allowbreak-sheet} & Draft a learning sheet from textbook sections. & \eff{15mm}{0mm}\\
    \texttt{write\allowbreak-tests} & Generate the test and validation sheet from a finalized learning sheet. & \eff{15mm}{0mm}\\
    \multicolumn{3}{@{}l}{\cellcolor{groupgray}\textit{Review (AI reviewer)}}\\
    \texttt{review\allowbreak-learning\allowbreak-sheet} & Critique a sheet for motivation, intuition, examples, proof correctness. & \eff{15mm}{0mm}\\
    \texttt{review\allowbreak-tests} & Check scope alignment, correctness, consistency. & \eff{15mm}{0mm}\\
    \multicolumn{3}{@{}l}{\cellcolor{groupgray}\textit{Instructor-guided revision}}\\
    \texttt{revise} & Walk the instructor through a sheet chunk by chunk; audit tests. & \eff{6.75mm}{8.25mm}\\
    \multicolumn{3}{@{}l}{\cellcolor{groupgray}\textit{Grading}}\\
    \texttt{grade\allowbreak-homework} & Grade a folder of submissions vs.\ solutions; flag uncertain items. & \eff{12.75mm}{2.25mm}\\
    \texttt{homework\allowbreak-report} & Build a per-student report with marks, feedback, flags. & \eff{15mm}{0mm}\\
    \bottomrule
  \end{tabular}
\end{table}
No AI draft reaches students without instructor review. This matters because
student acceptance of a strict gate depends on trust in the weekly materials.
An unclear learning sheet wastes preparation time. A drifting validation sheet
makes the test unfair. An AI-written proof with a hidden mistake teaches the
mistake. The harness was designed to catch these problems before students see
the materials and to keep marking fast enough that feedback shapes the next
week.

\section{Context and method}
\label{sec:method}

\textbf{Course and students.} DSAA 3071 is an upper-level course in theory of
computation at HKUST(GZ), the study of formal models of computing: what machines
can and cannot solve, and how to prove the difference. The Spring 2026 run
lasted 13 weeks and had 18 end-of-term survey respondents, most of them
third-year data-science students. Before the course,
56\% reported using AI every day and 83\% had at least informal prior exposure
to the topics. The subject is a useful test case because it is easy to feel
that one understands an explanation while still being unable to write a proof.

\textbf{Weekly rhythm.} Each week used three hours: a 30-minute closed-book
test on the previous week's material, a 120-minute self-paced study block, and
a 20-minute discussion. The new validation sheet was released right after the
test, while the gap was fresh; its answers followed two days later. During
study, students worked in groups of about four while the instructor and TAs
circulated and surfaced confusion. Widespread misconceptions triggered a short
whole-room explanation. Assessment was intentionally simple: 30\% weekly
in-class tests and 70\% final exam, with no homework, projects, or
presentations, again for signal quality (Section~\ref{sec:intro}).

Table~\ref{tab:time} gives \emph{estimated} recurring weekly time inputs for a
later offering after reusable materials exist. These are planning estimates, not
logged measurements, and one-shot setup work is excluded. The instructor
estimate reflects AI handling drafting, review, and first-pass marking.

\begin{table}[t]
  \caption{Estimated recurring weekly time inputs (planning estimates, not
  logged). Student rows count centralized or assigned time and exclude the
  self-paced block; staff rows exclude reusable one-shot work.}
  \label{tab:time}
  \small
  \begin{tabular}{@{}lcc@{}}
    \toprule
    \textbf{Weekly time input} & \textbf{Traditional} & \textbf{TDAA (later run)} \\
    \midrule
    \multicolumn{3}{@{}l}{\cellcolor{groupgray}\textit{Student centralized/assigned time}}\\
    Lecture & 3\,h & 0\,h \\
    Closed-book test & 0\,h & 30\,min \\
    Discussion & 0\,h & 20\,min \\
    Assigned homework & 1\,h & 0\,h \\
    \textbf{Student total} & \textbf{4\,h/wk} & \textbf{50\,min/wk} \\
    \multicolumn{3}{@{}l}{\cellcolor{groupgray}\textit{Recurring staff time}}\\
    Instructor total & 7--10\,h/wk & 4--5\,h/wk \\
    TA total & 3--6\,h/wk & 3\,h/wk \\
    \bottomrule
  \end{tabular}
\end{table}

\textbf{Evidence.} The reflection in Section~\ref{sec:reflection} draws on three
sources already in the project repository: (1) an end-of-term survey
($N=18$) with agreement items, component rankings, and free-text comments; (2)
the aggregate score record for all twelve weekly tests (weekly $n=18$--$24$ by
attendance); and (3) a git-history audit of later semantic repairs to the
generated answer keys. The survey was voluntary and anonymous, collected at the
end of term; it carries the usual self-report and self-selection limits noted in
Section~\ref{sec:threats}. We make no causal or learning-gain claims: the
analysis is about acceptance and operational feasibility, not measured
effectiveness.

\section{Reflection on the run}
\label{sec:reflection}

\subsection{RQ1: Acceptance of the gate}

We treat acceptance as a design mechanism: did students experience the strict
weekly gate as useful and attainable when given an explicit path to prepare?

\begin{figure*}[t]
  \centering
  \includegraphics[width=\textwidth]{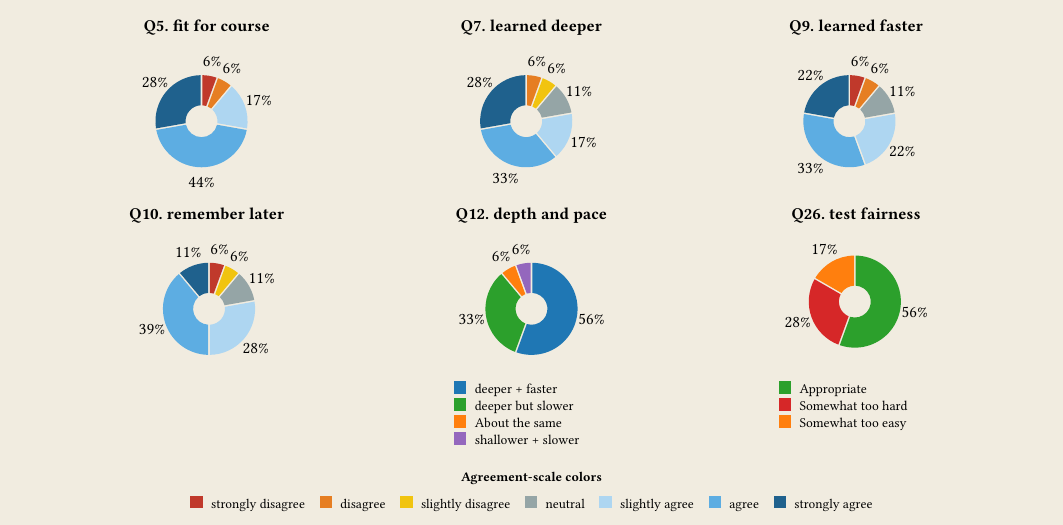}
  \caption{Student survey overview ($N=18$). The first four pies use the same
  seven-point agreement scale (1\,=\,strongly disagree, 7\,=\,strongly agree);
  the last two show the deeper-vs-faster comparison with a traditional lecture
  (Q12) and perceived test difficulty (Q26). Percentages are rounded from counts
  out of 18.}
  \label{fig:survey}
\end{figure*}

\textbf{Self-reported acceptance.} On seven-point items (max $=7$), students
averaged 5.61 on whether the format fit the subject, 5.50 on whether they
learned more deeply, 5.28 on whether they learned faster, and 5.17 on whether
they expected to remember the material after a year. Comparing with a
traditional lecture course, 56\% said it was both deeper and faster and a
further 33\% said deeper but slower. Figure~\ref{fig:survey} shows these
distributions with the test-difficulty result.

Free-text comments illuminate why. Students mentioned learning at their own
pace, asking questions immediately when confused, and exploring beyond a fixed
lecture path. One wrote that the format let them ``control your learning speed
to better understand.''

\textbf{Test accountability.} The weekly closed-book test gave the self-paced
block its weight. Asked which components they would miss most if removed,
students ranked the self-paced+AI block first (mean rank 2.56) and the
30-minute closed-book test second (3.22), ahead of mini-lectures, floor-walking,
same-session grading, and recap. The comments make the mechanism plain: ``Weekly
exams push me to study''; another credited ``having validation set to practice
and test set to do self-check''; a third said weekly tests kept them from
leaving everything to the final.

\begin{figure*}[t]
  \centering
  \includegraphics[width=\textwidth]{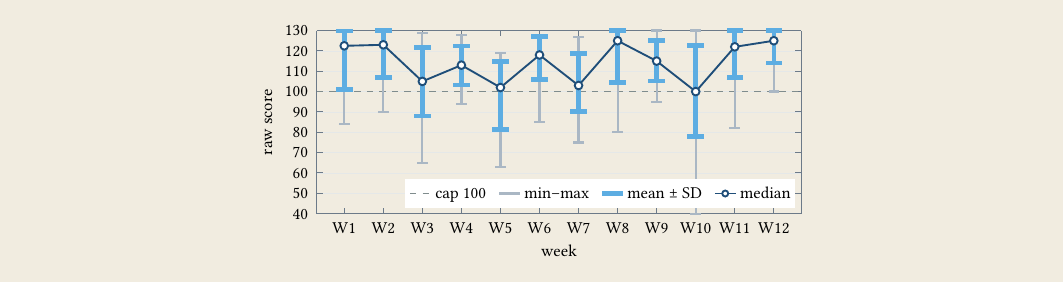}
  \caption{Weekly test score distribution (raw points out of 130; dashed line
  marks the 100-point credit cap; weekly $n=18$--$24$). Dark dots/line are
  weekly medians, thick bars mean\,$\pm$\,SD, thin bars min--max.}
  \label{fig:grades}
\end{figure*}

\textbf{Attainable pressure.} The raw-score distribution
(Figure~\ref{fig:grades}) reinforces this: every weekly median reached at least
the 100-point credit cap; even the hardest week by median hit exactly 100.
Across all 251 recorded student-test observations, 201 (80.1\%) were at least
100 raw points. The gate was real but attainable for most students. The survey
agrees: 56\% judged the weekly tests appropriate in difficulty, 28\% somewhat
too hard, and 17\% somewhat too easy.

\subsection{The teacher's role moved, not vanished}

The teacher's classroom role shifted from routine exposition toward diagnosis
and coaching. Mini-lectures ranked fourth of seven components (mean rank 3.94)
and floor-walking fifth (4.28). Neither was top-ranked, but both were still
wanted: 7 of 18 students asked for short recorded videos on the hardest
concepts, and 5 of 18 for more mini-lectures on hard topics. Self-regulation items explain why this role
cannot be handed to AI. Nine of 18 said they rarely or only sometimes thought
first before asking AI. Seven of 18 said they rarely or only sometimes could
tell when an AI explanation was wrong. As one student put it, ``the difficult ones should
be clearly explained by teacher.'' This supports a targeted teacher
role: diagnosing confusion, marking crucial ideas, and explaining hard concepts
when self-paced work is insufficient. It does not support a stronger causal
claim, since the run did not log every intervention.

\subsection{RQ2: Production feasibility}

The repository holds the 36 weekly student-facing files counted in
Section~\ref{sec:harness}, plus 9 project-specific skills.

\textbf{Learning materials.} AI-generated learning sheets were usable at course
scale, but only inside a review harness. Twelve of 18 respondents agreed or
strongly agreed the sheets balanced intuition and rigor. Local criticism asked
for more examples and clearer proof steps. One week's draft contained a plausible but
wrong proof shortcut; the reviewer flagged it and the revision used a safer
argument. Better agents should reduce such failures, but the result does not
justify removing human review.

\textbf{Tests and validation.} A git-history audit of the current test and
validation files counted 273 distinct answer-key question blocks. Subsequent
commits contained 10 semantic repairs (proof logic, missing conditions, theorem
references, or analysis), leaving 263 of 273 blocks (96.3\%) with no post-release
repair. This is not a controlled benchmark, but it is operational evidence that
agents can produce mostly correct formal answer keys at course scale. The
remaining human role is calibration: choosing which questions to keep,
controlling difficulty, and polishing ambiguous wording.

\textbf{Marking.} AI-assisted marking became practical because multimodal agents
can read scanned submissions, compare them against a provided solution, and
return structured marks with uncertainty flags. Consistent with the grading
literature (Section~\ref{sec:background}), the harness routes uncertain items to
the instructor rather than guessing. Grading is thus an automation skill with an
advisor-style escalation (Section~\ref{sec:harness}). Because the
100-point cap sits below the 130-point ceiling, small marking uncertainty often
has no effect on credited score. The run supports an operational claim: AI
marking can absorb routine work while preserving human review for doubtful
cases.

\textbf{Skills as captured judgment.} The instructor was running a feedback
system, not just producing documents. When tests were miscalibrated, the
test-writing skill was adjusted. When students missed crucial ideas, the
learning-sheet reviewer was changed to foreground them. When answer keys failed
in recurring ways, those checks were added to the grading skill. The git history
records this evolution across many \texttt{fix} and \texttt{revise} commits
clustering into four checks: textbook alignment, mathematical correctness,
weekly scope, and pedagogical order. Skill optimization did not eliminate
teacher judgment; it made teacher judgment reusable.

\subsection{RQ3: Where the design strained}

\textbf{Hard-topic scaffolding was insufficient.} Free-text complaints
concentrate on later proof-heavy topics: reductions, NP-completeness, PSPACE,
and an overloaded week. Students described these as ``too terse'' and asked for
more figures and examples, fitting the cognitive-load and proof-scaffolding work
in Section~\ref{sec:background}. The design response is
not to abandon the gate but to make the harness detect high-friction topics and
require extra scaffolding there.

\textbf{Validation was both anchor and trap.} Students credited the validation
sheet with making tests navigable, but the same mechanism can narrow learning.
One worried about relying on the validation set instead of true understanding.
Another said it could still ``surprise'' them. If validation is too close to the
test, students pattern-match; if too far, the gate feels unfair. The harness
began trimming near-duplicate items and adding redundancy checks across the test
and validation. The next version should make the separation systematic via an
explicit scope map.

\textbf{Improvement signals pointed to artifacts, not format.} Only 4 of 18
selected fewer tests as a top improvement. The strongest signals were 11 of 18
for more worked examples, 11 for a practice-problem bank, 9 for an AI-workflow
guide, and 7 for short videos. The next iteration should improve artifacts and
targeted teacher intervention before weakening the weekly gate.

\section{Limitations and threats to validity}
\label{sec:threats}

This is a single-course experience report, and the evidence is bounded
accordingly. The survey has $N=18$ and is self-reported and self-selected; it
measures perception, not learning. There is no control group and no pre/post
instrument, so no causal or effectiveness claim is made. Per-test scores are
not linked across weeks, and there is no final-exam outcome data, so we cannot
relate weekly performance to final attainment. The workload figures in Table~\ref{tab:time} are planning estimates,
not time logs. The git-history audit uses an author-defined notion of a
``semantic repair'' and is descriptive, not a benchmark. The instructor was also
an author, which can bias material design and interpretation. Finally, the
subject is small and proof-heavy; transfer to large or non-proof courses is a
hypothesis for future offerings, not a result.

\section{Adoption and conclusion}
\label{sec:conclusion}

\textbf{Adoption.} The observable parts of TDAA are reproducible: clear weekly
materials, self-paced study, strict closed-book checks, fast marking, and review
before students see AI-written material. We release the reusable harness as a
public starter template, separate from the course-specific content. It bundles
the document and website templates, the build commands, a GitHub Pages workflow,
and the named AI-assisted skills. With these, another instructor can bootstrap a
course, generate and review a week, build the site, and run AI-assisted grading. A full
step-by-step recipe and the practical adoption guide ship inside the harness
itself.\footnote{Reusable harness and companion artifact:
\url{https://github.com/GiggleLiu/TDAA-Go}.} The
structural requirement is what matters most: without a harness, twelve weekly
bundles and twelve graded tests are not realistic for one person.

\textbf{Conclusion.} We design TDAA, a test-driven, AI-assisted course model
built on one pairing. A weekly closed-book gate makes lecture-free, self-directed
study accountable, and a human-reviewed AI harness makes that gate sustainable
for one instructor. One thirteen-week run suggests the design is workable, though
not yet proven. Students accepted the gate, treating it as useful accountability
rather than punishment, and asked for better examples rather than fewer tests.
One instructor sustained twelve cycles of drafting, testing, and grading because
agents absorbed the repetitive volume while human review caught their errors. The
strain showed where the topics were hardest: later proof-heavy material, where
scaffolding and practice-to-test calibration still need work. The teacher's role
moved rather than vanished, from lecturing to designing the path, overseeing
AI-made materials, and diagnosing the confusion the materials left behind. Given
the limits in Section~\ref{sec:threats}, we offer it as a reusable design
pattern, not a causal claim.

\bibliographystyle{ACM-Reference-Format}
\bibliography{refs-arxiv}

\end{document}